\documentclass[twocolumn,showpacs,preprintnumbers,superscriptaddress,amsmath,amssymb]{revtex4-1}
\usepackage{amssymb}		%math formula
\usepackage{amsmath}
\usepackage{graphicx}
\usepackage[normalem]{ulem}
\usepackage{multirow}
\usepackage{appendix}
\usepackage{CJK}
\usepackage[usenames]{color}
\usepackage{bm}
\usepackage[colorlinks,linkcolor=blue,urlcolor=blue,citecolor=blue]{hyperref}
\usepackage{booktabs} 	% table!
\usepackage{leftidx}

\usepackage{soul,xcolor}
\setstcolor{red}

\usepackage[all]{hypcap} % cross-reference

\makeatletter

\newcommand{\Rmnum}[1]{\expandafter\@slowromancap\romannumeral #1@}
\makeatother

\newcommand\redsout{\bgroup\markoverwith{\textcolor{red}{\rule[0.5ex]{2pt}{0.4pt}}}\ULon}

\begin{document}
\begin{CJK*} {UTF8} {gbsn}
%\begin{CJK*} {GB} {gbsn}

\title{Collision centrality and system size dependences of light nuclei production via dynamical coalescence mechanism}

\date{\today}

\author{Yi-Lin Cheng(程艺琳)}
\affiliation{Shanghai Institute of Applied Physics, Chinese Academy of Sciences, Shanghai 201800, China}
\affiliation{Key Laboratory of Nuclear Physics and Ion-beam Application (MOE), Institute of Modern Physics, Fudan University, Shanghai 200433, China}
\affiliation{School of Nuclear Sciences and Technology, University of Chinese Academy of Sciences, Beijing 100049, China}

\author{Song Zhang(张松)} \thanks{Email: song\_zhang@fudan.edu.cn}
\affiliation{Key Laboratory of Nuclear Physics and Ion-beam Application (MOE), Institute of Modern Physics, Fudan University, Shanghai 200433, China}

\author{Yu-Gang Ma(马余刚)} \thanks{Email:  mayugang@fudan.edu.cn}
\affiliation{Key Laboratory of Nuclear Physics and Ion-beam Application (MOE), Institute of Modern Physics, Fudan University, Shanghai 200433, China}
%\affiliation{Shanghai Institute of Applied Physics, Chinese Academy of Sciences, Shanghai 201800, China}

\begin{abstract}
Light (anti-)nuclei  in relativistic heavy-ion collisions are considered to be formed by  the coalescence mechanism of (anti-)nucleons in the present work. Using a dynamical phase-space coalescence model coupled with a multi-phase transport (AMPT) model, we explore the formation of light clusters such as deuteron, triton and their anti-particles in different centralities for $^{197}$Au + $^{197}$Au collisions at $\sqrt{s_{NN}} = 39$ GeV. The calculated transverse momentum spectra of protons, deuterons, and tritons  are comparable to those of experimental data from the RHIC-STAR collaboration. Both coalescence parameters $B_{2}$ for (anti-)deuteron and $B_{3}$ for triton increase with the transverse momentum as well as the collision centrality,  and they are comparable  with the measured values in experiments. The effect of system size on the production of light nuclei is also investigated by $^{10}$B + $^{10}$B, $^{16}$O + $^{16}$O, $^{40}$Ca + $^{40}$Ca, and $^{197}$Au + $^{197}$Au systems in central collisions. The results show that yields of light nuclei increase with system size, while the values of coalescence parameters present an opposite trend. It is interesting to see that the system size, as well as  the centrality dependence of $B_A$ ($A$ = 2, 3), falls into the same group, which further demonstrates production probability of light nuclei is proportional to the size of the fireball. Furthermore, we  compare our coalescence results with other models, such as the thermal model and analytic coalescence model, it seems that the description of light nuclei production is consistent with each other.

\end{abstract}

\maketitle
\section{Introduction}

Quantum Chromodynamics (QCD) predicts that a new state of matter, namely the quark-gluon plasma (QGP), is likely to be formed  in an extremely high temperature or density
environment~\cite{QCD-QGP}, which could  be existed in the microseconds after the big bang. Studying this new matter is of great significance for us to have a comprehensive understanding not only of the basic composition and  interaction of matter but also the information of the early evolution of the universe. Relativistic heavy-ion collision
is currently considered as a unique way in the laboratory to detect such extremely high-temperature and -density QCD matter and then explore the QGP phase structure. However, the QGP state can only survive at a relatively short stage in the collision process, and it is soon hadronized
as the system's temperature and density rapidly decrease, then hadrons will interact with each other. While hadronic interaction ceases, the particle approaches a kinetic freeze-out stage. Experimentally one can infer the properties of the early QGP by exploring the kinetic freeze-out particles. Therefore, exploring the properties of QGP and QCD critical point from the regular hadronic matter to the QGP phase remains of great interest to the field~\cite{WongRHICIntro,Shuyak,PBM,CHEN20181,NSTLuoCQ,NSTSongFlow,FUKUSHIMA201399,Bzdak}.

Considering the light nuclei have small binding energy, it is also an open question of how they can survive from the hot nuclear matter. They might be disintegrated and regenerated through the coalescence of nucleons which are de-coupled from the hot and dense system, so the production of light nuclei can be used to extract the information of freeze-out nucleon distributions~\cite{STARdeuteronPRC2019} and to understand how the QGP expands, cools and hadronizes. These pieces of information provide crucial insights for dynamical mechanism and space-time evolution of heavy-ion collisions~\cite{Lao2018,Tang,CPL2,Wang2019,Huang2020}.
Recently the light nucleus also demonstrates its significance to search for the possible critical point in the phase diagram of strongly interacting quark
matter~\cite{SUN2018499,Yu2020,DengMa2020,Luo2020,SUN2017103,ShuryakPhysRevC2019}.
Theoretical study about the light clusters has been undertaken for a long time and several models or methods are used to explore the production of light nuclei. Thermal models~\cite{2020EPJA...56..241S,2020arXiv201103826A,PhysRevC.101.034906,PhysRevC.79.034909} have successfully described the yields of hadrons and nuclei. Besides, the coalescence model has been used to describe the production of light nuclei for many
years~\cite{PhysRevC.55.1443,Yan2006,ZHANG2010224,CHO2017279,WangTT2019,PhysRevC.102.044912,SUN2019132}. These calculations by using a similar coalescence mechanism coupled with phase-space distribution from different models, such as blast-wave model and transport model, seem to resemble each other of description for light nuclei production at RHIC and LHC energies. The production of light nuclei can be also described by the kinetic equations~\cite{DANIELEWICZ1991712,Oliinychenko2019}. Especially recently, the relativistic kinetic equations with their nonlocal collision integrals were also solved for successfully describing light (anti-)nuclei production from the many-body scatterings in high-energy nuclear collisions~\cite{Sun2021}.

In the present work, the system size dependence (centrality and collision system) are payed more attention. We investigate the production of deuteron and triton in relativistic heavy-ion collision by means of A Multi-Phase Transport (AMPT) model~\cite{AMPT2005}  followed by a dynamical coalescence model for $^{197}$Au + $^{197}$Au collisions at different centralities as well as for the central collisions of  $^{10}$B + $^{10}$B, $^{16}$O + $^{16}$O, and $^{40}$Ca + $^{40}$Ca  at  $\sqrt{s_{NN}} = 39$ GeV.
The coalescence factor extracted from the transverse momentum spectra of light nuclei and proton represents the coalescence probability, and it is related to the source volume that decreases with the increasing of constituent momentum of coalesced  nucleons~\cite{PhysRevC.99.054905}. The transverse momentum ($p_T$) distribution and the coalescence parameters ($B_A$) of light nuclei are comparable to the experimental data.
On the other hand, the properties of QGP are sensitive to the initial geometry and the dynamical fluctuations in heavy-ion collisions, and the system size scan experiment has been proposed at RHIC energies recently~\cite{SHuang2020sysScan}. These experiments will provide us more information of the initial geometry distribution and fluctuation effects on momentum distribution at the final stage, and some related theoretical analytical works have been performed~\cite{PRC2029sysScanLHC,PRL2014He3Au,smallSystem-th-SHL2019-1,PhysRevC.102.041901,ZHANG2020135366,PhysRevC.101.034906,Liu2017}.
Along this direction, a system scan of the coalescence parameters is undertaken in the present work and it is found that $B_A$  falls into the same group for its centrality dependence when both the system size and centrality are expressed by charged particle multiplicity（$\left<N_{ch}\right>$)，including $\pi^{\pm}$, $k^{\pm}$, $p$, $\bar{p}$, which indicates that light nuclei production essentially depends on the size of the fireball.

The paper is arranged as follows: In Section II, a brief description of the AMPT model which is used to generate the nucleon phase-space distribution at the freeze-out stage is presented. Also,the coalescence model for the light cluster is described, including the Wigner phase space density functions for the (anti)deuteron and (anti)triton. In section III, the results of $p_T$ distribution and the coalescence parameters of (anti)deuteron and triton from $^{10}$B + $^{10}$B, $^{16}$O + $^{16}$O and $^{40}$Ca + $^{40}$Ca in central collisions as well as
$^{197}$Au + $^{197}$Au collisions at different centralities are compared to the available experimental data. Finally, a conclusion is presented in Section IV.

\section{MODEL and ALGORITHMS}
\subsection{AMPT model}

\begin{figure}[htbp]
	\centering
	\includegraphics[width=8.5cm]{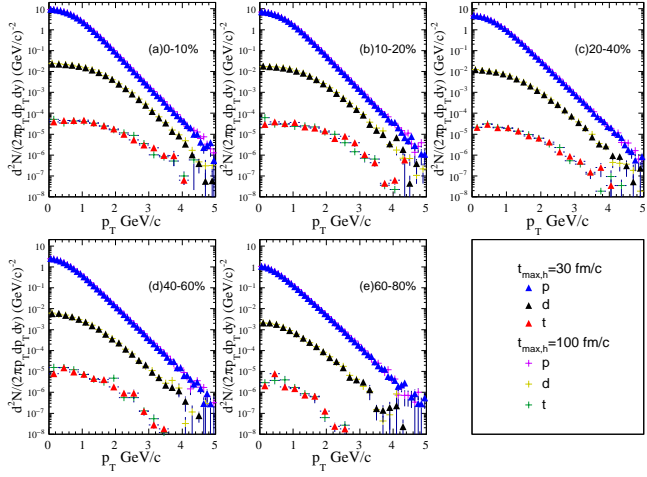}
	\caption {Transverse momentum $p_T$ spectra at mid-rapidity $(|y| < 0.5)$  of proton, deuteron and triton in $^{197}$Au + $^{197}$Au collisions for different centralities at $\sqrt{s_{NN}} = 39$ GeV with  the maximum hadronic rescattering time of 30 $fm/c$ and 100 $fm/c$.}
	\label{fig:NTMAX500_AUAU}
\end{figure}

\begin{figure}[htbp]
	\centering
	\includegraphics[width=8.5cm]{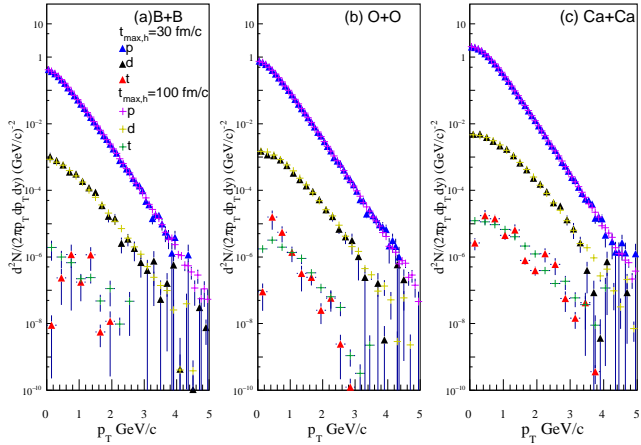}
	\caption {Transverse momentum $p_T$ spectra at mid-rapidity $(|y| < 0.5)$ for 0$-$10\% central collisions of $^{10}$B + $^{10}$B, $^{16}$O + $^{16}$O, $^{40}$Ca + $^{40}$Ca systems  at $\sqrt{s_{NN}} = 39$ GeV with the maximum hadronic rescattering time of 30 $fm/c$ and 100 $fm/c$.
}
	\label{fig:dif-NTMAX500}
\end{figure}

A multi-phase transport model ~\cite{AMPT2005} was used to provide the phase-space of nucleons in this work. The model is composed of four parts: the HIJING model~\cite{HIJING-1,HIJING-2} is used to simulate the initial conditions, the Zhang's Parton Cascade (ZPC) model~\cite{ZPCModel} is employed to describe partonic interaction, the Lund string fragmentation or coalescence model is used for the hadronization process, and A Relativistic Transport (ART) model~\cite{ARTModel} is applied to describe the hadronic rescattering process. As an event generator used in this work, the AMPT model
outputs the phase-space distribution at the final stage in the hadronic rescattering process (ART model~\cite{ARTModel}) with considering baryon-baryon, baryon-meson, and meson-meson elastic and inelastic scatterings, as well as resonance decay or week decay. In Refs.~\cite{ARTModel,AMPT2005} the interaction cross section was presented and extended. The hadronic rescattering time would affect  light nuclei spectra and yield which are based on the  phase-space information of nucleons from the AMPT model. Refs.~\cite{AMPT2005,PhysRevC.93.054911} suggest the maximum hadronic rescattering time ($t_{max,h}$), which means to cease a hadron interacting with others if it still dose not reach freeze-out state at that time, 30 $fm/c$ for the RHIC energy region and 200 $fm/c$ for the LHC energy region. Here  the $p_T$ spectra of $p$, $d$ and $t$ with the cutoff of the maximum hadronic rescattering time of 30 $fm/c$ and 100 $fm/c$ are checked.  Fig.~\ref{fig:NTMAX500_AUAU} shows the $p_T$ spectra  of proton, deuteron and triton of  $^{197}$Au + $^{197}$Au collisions at mid-rapidity $(|y| < 0.5)$  for different centralities at $\sqrt{s_{NN}} = 39$ GeV, and Fig.~\ref{fig:dif-NTMAX500} presents the  $p_T$  results for the 0$-$10\% central collisions of $^{10}$B + $^{10}$B, $^{16}$O + $^{16}$O, $^{40}$Ca + $^{40}$Ca at $\sqrt{s_{NN}} = 39$ GeV and mid-rapidity $(|y| < 0.5)$. We can find that these two cases are very  close to each other. Afterwords we choose the case of $t_{max,h}$ = 100 $fm/c$ for the following calculations. We would mention that the AMPT model has been successfully used to simulate physics in heavy-ion collisions at the RHIC and LHC energies~\cite{AMPT2005,PhysRevC.93.054911,AMPT2021} and the detailed parameter configurations can be found therein.

\subsection{Dynamical coalescence model}

In the coalescence model~\cite{CSERNAI1986223}, the invariant yields of light nuclei with charge number $Z$ and atomic mass number $A$ can be described by the yields of cluster constituents (protons and neutrons) multiplying by a coalescence parameter $B_{A}$,
\begin{eqnarray}%{equation}
\begin{aligned}
E_{A}\frac{d^{3}N_{A}}{dp^{3}_{A}}\!&=\!B_{A}(E_{p}\frac{d^{3}N_{p}}{dp^{3}_{p}})^{Z}(E_{n}\frac{d^{3}N_{n}}{dp^{3}_{n}})^{A\!-\!Z}\\
&\!\approx\! B_{A}(E_{p}\frac{d^{3}N_{p}}{dp^{3}_{p}})^{A},
\label{BA}
\end{aligned}
\end{eqnarray}%{equation}
where $p_p$ and  $p_n$ are the momenta of proton and neutron, respectively, and $p_{A}$ is the momentum of the nucleus with the mass number $A$ which is approximate $A$ times of proton momentum, i.e. $A p_{p}$, assuming that the distributions of neutrons and protons are the same. The coalescence parameter $B_{A}$ related to the local nucleon density reflects the probability of nucleon coalescence. The coalescence parameter $B_A$ is also related to the effective volume of the nuclear matter at the time of coalescence of nucleons into light nuclei, called nucleon correlation volume $V_{eff}$ \cite{CSERNAI1986223}, i.e. $B_A \propto  1/V_{eff}^{A-1}$.

The dynamical coalescence model can give the probability of light nuclei ($M$-nucleon cluster) by the overlap of the cluster Wigner phase-space density with the nucleon phase-space distributions at  an equal time in the $M$-nucleon rest frame at the freeze-out stage~\cite{CHEN2003809}. The momentum distribution of a cluster in a system containing $A$ nucleons can be expressed by,
\begin{small}
\begin{eqnarray}%{equation}
\begin{aligned}
\frac{d^{3}N_{M}}{d^{3}K}&=G\begin{pmatrix} A\\\\ M\\\\\end{pmatrix}\begin{pmatrix}M\\\\Z\\\\\end{pmatrix} \frac{1}{A^{M}}\!\int\!\left[\prod_{i=1}^{Z}f_{p}(\vec{r}_{i},\vec{k}_{i})\right]\\
&\left[\prod_{i=Z+1}^{M}f_{n}(\vec{r}_{i},\vec{k}_{i})\right]\times \rho^{W}(\vec{r}_{i_{1}},\vec{k}_{i_{1}},\cdots,\vec{r}_{i_{M-1}},\vec{k}_{i_{M-1}})\\
&\times\delta(\vec{K}-(\vec{k_{1}}+\cdots+\vec{k_{M}}))d\vec{r}_{1}d\vec{k}_{1}\cdots d\vec{r}_{M}d\vec{k}_{M},
\label{dynamicalCoal}
\end{aligned}
\end{eqnarray}%{equation}
\end{small}
where $M$ and $Z$ are the number of the nucleon and proton of the cluster, respectively; $f_{n}$ and  $f_{p}$ are the neutron and proton phase-space distribution functions at freeze-out, respectively; $\rho^{W}$ is the Wigner density function; $\vec{r}_{i_{1}},\cdots,\vec{r}_{i_{M-1}}$ and $\vec{k}_{i_{1}},\cdots,\vec{k}_{i_{M-1}}$ are the relative coordinates and momentum in the $M$-nucleon rest frame; the spin-isospin statistical factor $G$ is 3/8 for deuteron and 1/3 for triton~\cite{CHEN2003809}, note whether to consider the isospin effect is still an unresolved problem, neglecting the isospin effect can be found in~\cite{PhysRevC.98.054905,PhysRevC.103.064909}. While the neutron and proton phase-space distribution comes from the transport model simulations, the multiplicity of a $M$-nucleon cluster is then given by,
\begin{small}
\begin{eqnarray}%{equation}
\begin{aligned}
N_{M} = G\!\int\!\sum_{i_{1}>i_{2}>\cdots>i_{M}}d\vec{r}_{i_{1}}d\vec{k}_{i_{1}}\cdots d\vec{r}_{i_{M-1}}d\vec{k}_{i_{M-1}}\\
\langle \rho^{W}_{i}(\vec{r}_{i_{1}},\vec{k}_{i_{1}},\cdots,\vec{r}_{i_{M-1}},\vec{k}_{i_{M-1}})\rangle,
\label{multiplicity}
\end{aligned}
\end{eqnarray}%{equation}
\end{small}
where the $\left<\cdots\right>$ denotes the event averaging.

%%%%%%%%szhang
\subsection{Wigner phase-space density}

The Wigner phase-space density of (anti)deuteron is assumed as~\cite{CHEN2003809},
\begin{small}
\begin{eqnarray}%{equation}
\begin{aligned}
\rho^{W}_{d}(\vec{r},\vec{k}) = 8\sum^{15}_{i=1}c^{2}_{i}\exp\left(-2\omega_{i}r^{2}-\frac{k^{2}}{2\omega_{i}}\right)\\
+16\sum^{15}_{i>j}c_{i}c_{j}\left(\frac{4\omega_{i}\omega_{j}}{(\omega_{i}+\omega_{j})^{2}}\right)^{\frac{3}{4}}\exp\left(-\frac{4\omega_{i}\omega_{j}}{\omega_{i}+\omega_{j}}r^{2}\right)\\
\times \exp\left(-\frac{k^{2}}{\omega_{i}+\omega_{j}}\right)\cos\left(2\frac{\omega_{i}-\omega_{j}}{\omega_{i}+\omega_{j}}\vec{r}\cdot\vec{k}\right),
\end{aligned}
\label{wignerdeuteron}
\end{eqnarray}%{equation}
\end{small}
where the Gaussian fit coefficient $c_{i}$ and  $w_{i}$ are given in~Ref.~\cite{CHEN2003809}.
$\vec{k}$ = ($\vec{k}_{1}$-$\vec{k}_{2}$)/2 is the relative momentum  and $\vec{r}$ = ($\vec{r}_{1}$-$\vec{r}_{2}$) is the relative coordinate of proton and neutron inside deuteron.

The Wigner phase-space density of triton is obtained from a spherical harmonic oscillator~\cite{CHEN2003809,ZHANG2010224,SUN2015272},
\begin{small}
\begin{eqnarray}%{equation}
\begin{aligned}
\rho^{W}_{t}(\rho,\lambda,\vec{k}_{\rho},\vec{k}_{\lambda})\!=\!\int\psi(\rho\!+\!\frac{\vec{R}_{1}}{2},\lambda\!+\!\frac{\vec{R}_{2}}{2}\!)\!\psi^{*}\!(\!\rho\!-\!\frac{\vec{R}_{1}}{2},\lambda\!-\!\frac{\vec{R}_{2}}{2})\\
\times \exp(-i\vec{k}_{\rho}\cdot\vec{R}_{1})\exp(-i\vec{k}_{\lambda}\cdot\vec{R}_{2})3^{\frac{3}{2}}d\vec{R}_{1}d\vec{R}_{2}\\
=8^{2}\exp(-\frac{\rho^{2}+\lambda^{2}}{b^{2}})\exp(-(\vec{k}^{2}_{\rho}+\vec{k}^{2}_{\lambda})b^{2}),
\end{aligned}
\label{wignertriton}%{equation}
\end{eqnarray}%{equation}
\end{small}
where $\rho$ and $\lambda$ are relative coordinates, $\vec{k}_{\rho}$ and $\vec{k}_{\lambda}$ are the relative momenta in the Jacobi  coordinate, the parameter $b$ is obtained from the root-mean-square radius, 1.61 $fm$ for triton~\cite{CHEN2003809}.

In practice, the coalescence procedure by using Eq.~(\ref{multiplicity}) can not guarantee the energy conservation, such as for the formation of dueteron $p+n \rightarrow d$. If a proton and a neutron with momentum-energy $(\vec{k},E_p)$ and $(-\vec{k},E_n)$ coalesces a deuteron with $(\vec{0},m_d)$, and then the lost energy is $\Delta E = \sqrt{k^2+m_p^2} + \sqrt{k^2+m_n^2} - m_d$.
From Eq.~(\ref{wignerdeuteron}), it can be seen that the lost energy is ignorable since the Wigner density is suppressed exponentially at the large relative momentum. For the three-body case, a similar derivation can be obtained.  Actually, we made a numerical check for the effect of lost energy, it is found that it is negligible for the yield and spectra of the light nuclei production.

In this calculation, the AMPT model provides the phase-space of nucleons at the freeze-out stage in heavy-ion collisions and the followed coalescence model is coupled to give the transverse momentum $p_T$ spectra of deuterons ($d$) and tritons ($t$). Based on the obtained $p_T$ spectra, the yields of $d$ and $t$, as well as the coalescence parameters, are discussed.

\section{Results and Discussion}

\begin{figure}[htbp]
	\centering
	\includegraphics[width=8.5cm]{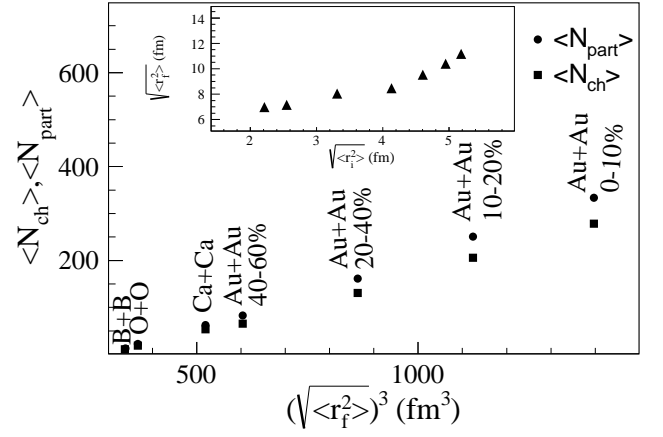}
	\caption {Relationship between $\left<N_{part}\right> $ ($\left<N_{ch}\right>$) and $\sqrt{\left<r_{f}^2\right>}$ to characterise the system size. The insert plots a correlation between  $\sqrt{\left<r_{i}^2\right>}$ and $\sqrt{\left<r_{f}^2\right>}$. }
	\label{fig:rRMS_if_Ntrack_Npart}
\end{figure}

To discuss the system size dependence of light nuclei production, some quantities of the collision systems are shown in figure~\ref{fig:rRMS_if_Ntrack_Npart}, such as $\left<N_{part}\right>$ representing the average number of participants, and  $\left<N_{ch}\right>$ denoting the average number of charged hadrons ($\pi^{\pm}$, $k^{\pm}$, $p$, $\bar{p}$) with a kinetic window of $0.2<p_T<2$ GeV$/c$ and rapidity $|y|<0.5$ (mid-rapidity), $\sqrt{\left<r_{i}^2\right>}$ representing the averaged radius of the initial collision zone which is calculated through the participants, $\sqrt{\left<r_{f}^2\right>}$ representing the averaged radius of the collision system at freeze-out stage which is calculated through the charged hadrons. It is seen that $\left<N_{part}\right>$ and  $\left<N_{ch}\right>$ are all proportional to collision system size at final stage, namely $\sqrt{\left<r_{f}^2\right>}$, for different collision systems. In the insert, the freeze-out radius of the collision system $\sqrt{\left<r_{f}^2\right>}$ increases with the initial radius of the collision zone, namely $\sqrt{\left<r_{i}^2\right>}$. So both  $\left<N_{part}\right>$ and  $\left<N_{ch}\right>$ can characterise the collision system size, and it is therefore convenient to discuss system size dependence of observables by comparing the  $\left<N_{ch}\right>$ -dependent results with experimental data in the following.

%fig0-rRMS_if_Ntrack_Npart

\subsection{$p_T$ spectra of $p$ ($\bar{p}$), $d$ ($\bar{d}$) and $t$ ($\bar{t}$)}

\begin{figure}[htbp]
	\includegraphics[width=8.3cm]{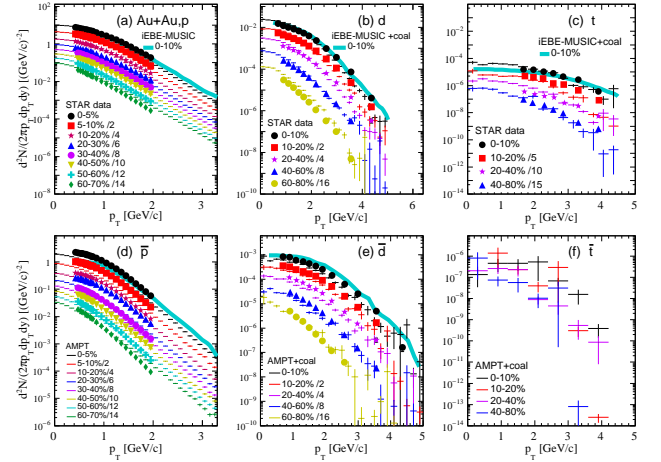}
    \caption {Transverse momentum $p_T$ spectra at mid-rapidity $(|y| < 0.5)$  of (anti)proton, (anti)deuteron and (anti)triton in $^{197}$Au + $^{197}$Au
collisions for different centralities at $\sqrt{s_{NN}} = 39$ GeV. Solid markers represent the experimental data from the STAR
collaboration~\cite{PRCSTAR2017SysBulk,STARdeuteronPRC2019,zhang2020light} and lines represent the model calculation results. The smooth lines represent the results of (anti)proton from the iEBE-MUSIC hybrid model, (anti)deuteron and triton from the iEBE-MUSIC hybrid model plus the coalescence model~\cite{PhysRevC.102.044912}.}
	\label{ptd-PT}
    \end{figure}

Figure~\ref{ptd-PT} presents the transverse momentum spectra of $p$ ($\bar{p}$), $d$ ($\bar{d}$) and $t$ ($\bar{t}$) calculated by the AMPT model coupling with the coalescence model in Au + Au collisions at $\sqrt{s_{NN}} = 39$ GeV. The results are shown for the collision centrality classes of 0$-$5\%, 5$-$10\%, 10$-$20\%, 20$-$30\%, 30$-$40\%, 40$-$50\%, 50$-$60\%, and 60$-$70\% for $p$ ($\bar{p}$) in
Fig.~\ref{ptd-PT}(a) and (d), 0$-$10\%, 10$-$20\%, 20$-$40\%, 40$-$60\%, and 60$-$80\% for $d$ ($\bar{d}$) in Fig.~\ref{ptd-PT} (b) and (e), 0$-$10\%, 10$-$20\%, 20$-$40\%, and 40$-$80\%
for $t$ ($\bar{t}$) in Fig.~\ref{ptd-PT}(c) and (f). It is found that the results can well describe the experimental data for $p$~\cite{PRCSTAR2017SysBulk}, $d$~\cite{STARdeuteronPRC2019} and $t$~\cite{zhang2020light} spectra from the STAR collaboration, especially in central collisions.
Besides, we compared  the transverse momentum spectra of $p$ ($\bar{p}$), $d$ ($\bar{d}$) and $t$ with the results from the iEBE-MUSIC hybrid model plus coalescence model~\cite{PhysRevC.102.044912}. Note that the isospin effect in the statistical factor in Eq.~(\ref{dynamicalCoal}) can result in a constant factor among the results~\cite{PhysRevC.102.044912,PhysRevC.103.064909} and does not affect the shape of the spectra. It is interesting to see that two models are consistent, which implies that the phase-space of the two models have similar properties or distributions.

\begin{figure}[htbp]
\centering
	\includegraphics[width=8.3cm]{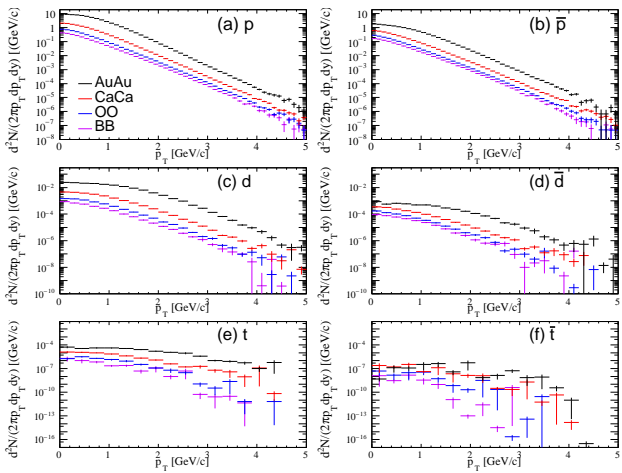}
	\caption {Transverse momentum $p_T$ spectra at mid-rapidity $(|y| < 0.5)$ of (anti)proton, (anti)deuteron and (anti)triton in 0$-$10\% central collisions of
$^{10}$B + $^{10}$B, $^{16}$O + $^{16}$O, $^{40}$Ca + $^{40}$Ca, and $^{197}$Au + $^{197}$Au at $\sqrt{s_{NN}} = 39$ GeV by using the AMPT model coupling with the coalescence model.}
	\label{all-pT-difsys}
\end{figure}

Figure~\ref{all-pT-difsys} shows the calculated transverse momentum spectra for $p$ ($\bar{p}$) ((a) and (b)), $d$ ($\bar{d}$)  ((c) and (d)) and $t$ ($\bar{t}$) ((e) and (f)) in $^{10}$B + $^{10}$B, $^{16}$O + $^{16}$O, $^{40}$Ca + $^{40}$Ca, and $^{197}$Au + $^{197}$Au in 0$-$10\% central collisions at $\sqrt{s_{NN}} $ = 39 GeV. The $p_{T}$ spectra present an obvious collision system dependence in central collisions and drop with the decreasing of the collision system size~\cite{ZHANG201176}.

\subsection{ $\langle dN/dy \rangle $ of $p$ ($\bar{p}$), $d$ ($\bar{d}$) and $t$ ($\bar{t}$)}

\begin{figure}[htbp]
	\centering
	\includegraphics[width=8.3cm]{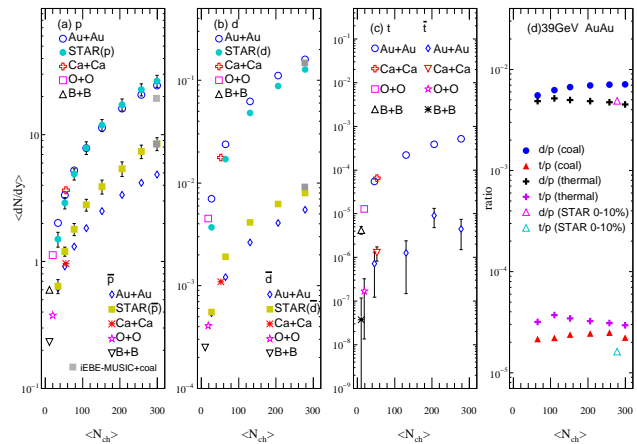}
	\caption {$\langle N_{ch} \rangle$ dependence of the yield $\langle dN/dy \rangle$ of $p$ ($\bar{p}$), $d$ ($\bar{d}$) and $t$ ($\bar{t}$) from  $^{10}$B + $^{10}$B, $^{16}$O + $^{16}$O, $^{40}$Ca + $^{40}$Ca, and $^{197}$Au + $^{197}$Au collision systems at 0$-$10\% centrality and  $\sqrt{s_{NN}}$ = 39 GeV are presented in (a)-(c). Results are compared with experimental data of $p$ ($\bar{p}$) and  $d$ ($\bar{d}$) in $^{197}$Au + $^{197}$Au collisions at $\sqrt{s_{NN}}$ = 39 GeV~\cite{PRCSTAR2017SysBulk,STARdeuteronPRC2019}. The gray markers are the results of the iEBE-MUSIC hybrid model plus coalescence model~\cite{PhysRevC.102.044912}. The comparison of $\langle dN/dy \rangle$ dependence of $d/p$ and $t/p$ between coalescence model and thermal model (fitted parameters is from Ref.~\cite{PRCSTAR2017SysBulk}) and the STAR data~\cite{PRCSTAR2017SysBulk,PhysRevC.103.064909} are shown in (d). }
	\label{yield-dif-AUAUall-BB-CaCa}
\end{figure}

The rapidity densities ($\langle dN/dy\rangle$) of $p$ ($\bar{p}$), $d$ ($\bar{d}$) and $t$ ($\bar{t}$) are calculated  in mid-rapidity as a function of  $\langle N_{ch}\rangle$ in $^{10}$B + $^{10}$B, $^{16}$O + $^{16}$O, $^{40}$Ca + $^{40}$Ca, and $^{197}$Au + $^{197}$Au collisions at $\sqrt{s_{NN}}$ = 39 GeV, as shown in  Fig.~\ref{yield-dif-AUAUall-BB-CaCa}.
It is found  that $\langle dN/dy\rangle$ of $p$ as a function of  $\langle N_{ch}\rangle$  (Fig.~\ref{yield-dif-AUAUall-BB-CaCa}(a)) can well describe the data~\cite{PRCSTAR2017SysBulk}  but underestimate $\bar{p}$ data in Au + Au collisions at $\sqrt{s_{NN}}$ = 39 GeV. For $d$ and $\bar{d}$ (Fig.~\ref{yield-dif-AUAUall-BB-CaCa}(b)), it presents the similar description quality to the data~\cite{STARdeuteronPRC2019}. $\langle dN/dy\rangle$ of $t$ and $\bar{t}$ as a function of  $\langle N_{ch}\rangle$  is presented in Fig.~\ref{yield-dif-AUAUall-BB-CaCa}(c). As shown in Fig.~\ref{yield-dif-AUAUall-BB-CaCa}(a) and (b), $\langle dN/dy\rangle$ of $p$ and $d$  are comparable to those from the iEBE-MUSIC hybrid model plus coalescence model~\cite{PhysRevC.102.044912} in central collisions, and little difference for anti-matter partners. In addition, the yields of these light (anti)nuclei for the 0-10\% central collisions of  $^{10}$B + $^{10}$B, $^{16}$O + $^{16}$O, and $^{40}$Ca + $^{40}$Ca systems at $\sqrt{s_{NN}}$ = 39 GeV are also shown in Fig.~\ref{yield-dif-AUAUall-BB-CaCa},
and it seems that they follow the similar $\langle N_{ch}\rangle$ systematics. In general, it is reasonably speculated that $\langle dN/dy \rangle $ of (anti)proton, (anti)deuteron and triton present an increasing trend with $\langle N_{ch}\rangle$ (collision system size) in different collision centralities as well as collision systems.

Furthermore, we calculate the $\langle N_{ch}\rangle$ dependence of ratios of $d/p$ and $t/p$ by using a thermal model~\cite{2004qgp3.book..491B},
\begin{eqnarray}%{equation}
\begin{aligned}
n_{i}(T,\vec{\mu}) \!&=\! \frac{\langle N_{i}\rangle}{V}\\
 &= \frac{Tg_{i}}{2\pi^{2}}\sum^{\infty}_{k=1}\frac{(\pm{1})^{k+1}}{k}\lambda^{k}_{i}m^{2}_{i}K_{2}(\frac{km_{i}}{T}),
\end{aligned}
\label{ni-thermal}
\end{eqnarray}%{equation}
where $\lambda_{i}(T,\vec{\mu}) =\exp(\frac{B_{i}\mu_{B}+S_{i}\mu_{s}+Q_{i}\mu_{Q}}{T})$, $B_{i}$, $S_{i}$ and $Q_{i}$ are the baryon number, strangeness number and charge number, $\mu_{B}$,  $\mu_{S}$ and $\mu_{Q}$, are their corresponding chemical potentials of particle $i$, $K_{2}$ is the modified Bessel function and the upper sign is for bosons and lower for fermions,
$g_{i}$ is the spin$-$isospin degeneracy factor. We use the parameters such as the chemical freeze-out temperature as well as the baryon chemical potential from Ref.~\cite{PRCSTAR2017SysBulk}. As shown in Fig.~\ref{yield-dif-AUAUall-BB-CaCa}(d), the  $d/p$ and $t/p$ ratios  of AMPT + coalescence model are bigger than  STAR data~\cite{PRCSTAR2017SysBulk,STARdeuteronPRC2019,DWZhang2019_dt}. And the $d/p$ ratio from the thermal model can describe the STAR data~\cite{PRCSTAR2017SysBulk,STARdeuteronPRC2019,DWZhang2019_dt} but overestimates the $t/p$ ratio, which is consistent with the results in references~\cite{DWZhang2019_dt,YuLuo2019_thermal,VOVCHENKO2020_thermal}.

\begin{figure}[htbp]
	\centering
	\includegraphics[width=8.6cm]{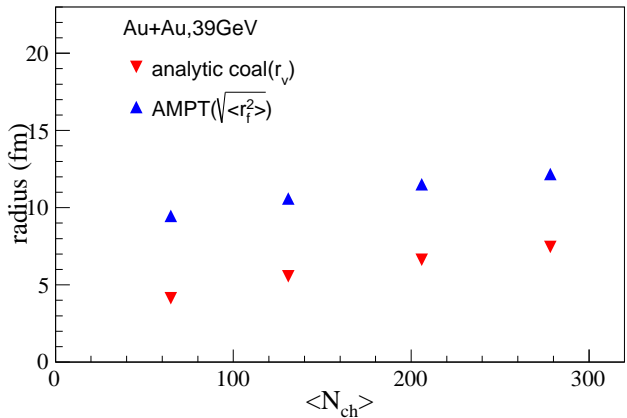}
	\caption {The comparison of  $\langle N_{ch} \rangle$ dependence of the fireball radius calculated directly from coordinates in the AMPT calculations as well as the results obtained by analytic coalescence model~\cite{PhysRevC.95.044905} in $^{197}$Au + $^{197}$Au collisions at $\sqrt{s_{NN}} = 39$ GeV.}
	\label{R_ana_coal}
\end{figure}

\begin{figure}[htbp]
	\centering
	\includegraphics[width=8.6cm]{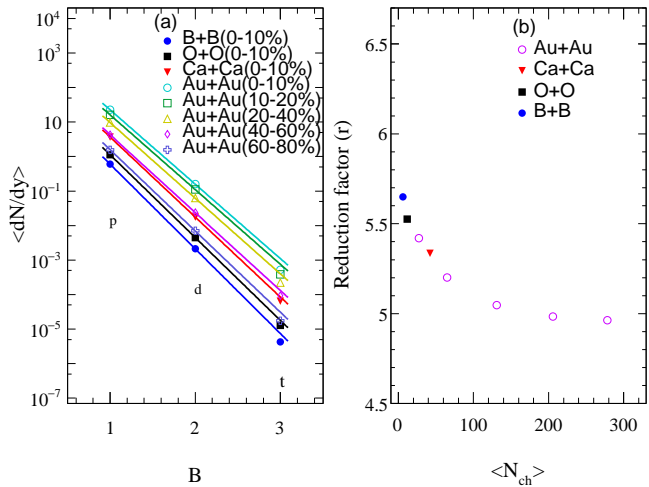}
	\caption {(a) The yields of proton, deuteron, and triton as a function of baryon number $B$ from the coalescence model for the  0$-$10\% $^{10}$B + $^{10}$B, $^{16}$O + $^{16}$O, and $^{40}$Ca + $^{40}$Ca collisions, as well as  for the 0$-$10\%, 10$-$20\%, 20$-$40\%, 40$-$60\%, and 60$-$80\% $^{197}$Au + $^{197}$Au collisions
at $\sqrt{s_{NN}} = 39$ GeV, (b) the extracted reduction factor $r$ obtained by fitting the yields of proton, deuteron, and triton with  a function of $N_0 e^{-rB}$ versus $\langle N_{ch} \rangle$.}
	\label{dndy-fit-ptd}
\end{figure}

 Figure~\ref{R_ana_coal} presents a comparison between $\langle N_{ch} \rangle$ dependence of the fireball radius ($\sqrt{\left<r^2_f\right>}$) calculated directly by the coordinates from the AMPT model and that $r_V$ from the  analytic coalescence model~\cite{PhysRevC.95.044905} in $^{197}$Au + $^{197}$Au collisions at $\sqrt{s_{NN}} = 39$ GeV. In analytic coalescence model~\cite{PhysRevC.95.044905}, a blast-wave-like parametrization is used for the phase-space configuration of constituent particles at freeze-out. We extract the effective volume by equation (25) in Ref.~\cite{PhysRevC.95.044905}, then the fireball radius $r_V$ can be calculated by assuming a spherical fireball.
We find that the both radii ($\sqrt{\left<r^2_f\right>}$ and $r_V$) present a similar  $\langle N_{ch} \rangle$ dependence, i.e. increasing as $\langle N_{ch} \rangle$. Of course, we notice that  the values of size  have model or calculation method dependence.

Figure~\ref{dndy-fit-ptd}(a) shows the $\langle dN/dy \rangle$ of proton, deuteron and triton as a function of baryon number $B$ from the coalescence model in 0$-$10\% $^{10}$B + $^{10}$B, $^{16}$O + $^{16}$O, and $^{40}$Ca + $^{40}$Ca collisions, as well as the 0$-$10\%, 10$-$20\%, 20$-$40\%, 40$-$60\%, and 60$-$80\% $^{197}$Au + $^{197}$Au collisions at $\sqrt{s_{NN}} = 39$ GeV.
The lines are the fits to the calculated results by a function of $N_0e^{-rB}$, here $N_0$ denotes amplitude, $B$ is the baryon number and $r$ is the reduction factor. It is found that the yields of proton, deuteron ,and triton in each collision system exhibit a decreasing exponential trend with the baryon number. The reduction factor~\cite{SHAH20166,SUN2015272}
 by fitting the  yields of proton, deuteron and triton as a function of $\langle N_{ch} \rangle$ is shown in Fig.~\ref{dndy-fit-ptd}(b). While the system size is expressed by  $\langle N_{ch} \rangle$, the reduction factor decreases sharply with the increasing of $\langle N_{ch} \rangle$ and then saturate at large $\langle N_{ch} \rangle$. This implies that light nuclei production becomes more difficult in small systems, especially for that with baryon number $B>$3 in the relativistic heavy-ion collisions.

\subsection{Coalescence parameters $B_{2}$ and $B_{3}$}

\begin{figure}[htbp]
	%\centering
	\includegraphics[width=8.8cm]{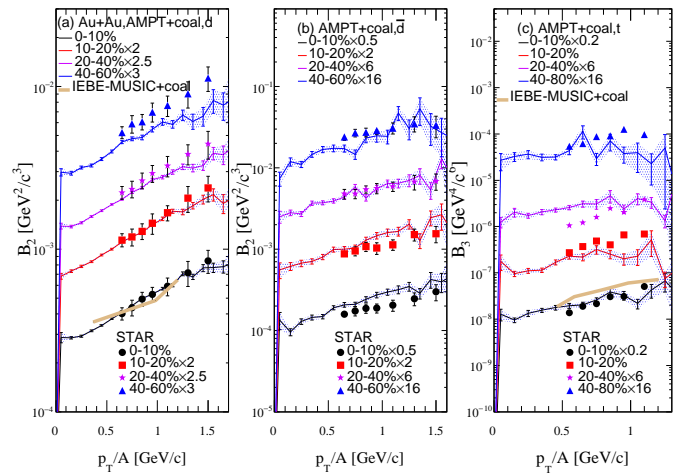}
	\caption {Coalescence parameters $B_{2}$ and $B_{3}$ as a function of $p_T/A$ for deuterons  (a),  anti-deuterons (b) and tritons  (c) for $^{197}$Au + $^{197}$Au
collisions at $\sqrt{s_{NN}}$ = 39 GeV at different centralities: 0$-$10\%, 10$-$20\%, 20$-$40\%, and 40$-$60\% (40$-$80\% for $t$). The solid markers are experimental data of (anti)deuterons and tritons from the STAR collaboration~\cite{STARdeuteronPRC2019}. The smooth lines are the results of $B_{2}$ and $B_{3}$  from the iEBE-MUSIC hybrid model plus coalescence model ~\cite{PhysRevC.102.044912}}.
	\label{B2}
\end{figure}

\begin{figure}[htbp]
	\centering
	\includegraphics[width=8.8cm]{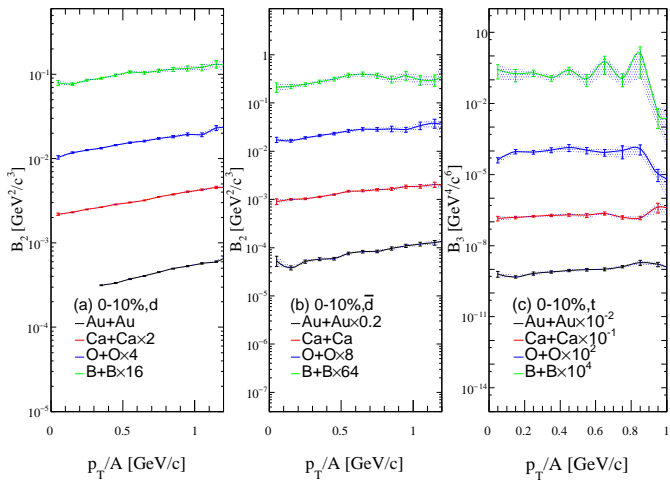}
	\caption {Coalescence parameters $B_{2}$ and $B_{3}$ as a function of $p_T/A$ for  deuterons  (a),  anti-deuterons (b)  and tritons  (c) from 0$-$10\% central collisions of  $^{10}$B + $^{10}$B, $^{16}$O + $^{16}$O, $^{40}$Ca + $^{40}$Ca, and $^{197}$Au + $^{197}$Au at $\sqrt{s_{NN}}$ = 39 GeV.
	}
	\label{B2-dif-PTA}
\end{figure}

To further characterize the system size dependence of light nuclei production, the coalescence probability of forming light clusters is investigated by the coalescence parameters $B_{A}$ ($A$ = 2 and 3) as defined in Eq.~(\ref{BA}). In panel (a) and (b) of Fig.~\ref{B2}, the calculated coalescence parameter $B_{2}$ are compared with the data measured by the STAR collaboration~\cite{STARdeuteronPRC2019}
in $^{197}$Au + $^{197}$Au collisions at RHIC energy of 39 GeV in 0$-$10\%, 10$-$20\%, 20$-$40\%, and 40$-$60\% (40$-$80\% for triton) centralities.
The calculated results present a similar trend with the experimental data，the coalescence parameters $B_{2}$ in panel (a), (b) and $B_{3}$ in panel (c) as a function of $p_{T}/A$ in different collision centralities always present an increasing trend, this might be due to the increasing correlation volume with the decreasing of $p_T$, leading to a higher coalescence probability for larger $p_T$ values.
In addition, the values of $B_{2}$ and $B_{3}$ decrease with collision centrality (i.e.  the more central collisions the less $B_{A}$), which suggests that source volume being larger in central collisions.
From the viewpoint of the coalescence probability of nucleons to form these light clusters, it is reasonable to have a bigger  coalescence probability while the distance between the protons and neutrons is smaller.
On the other hand, we note that the values of $B_{2}$ for deuterons are systematically larger than those of anti-deuterons in the same centrality, it is consistent with the experimental observation \cite{STARdeuteronPRC2019}, indicating that the correlated volume of baryons is smaller than  that of anti-baryons. Besides, the comparison of our results of $B_{2}$ and $B_{3}$  with the iEBE-MUSIC hybrid model plus coalescence model~\cite{PhysRevC.102.044912} is also shown in this figure, and the trend remains similar.

Furthermore, the coalescence parameter $B_{2}$ for (anti)deuterons as a function of $p_{T}/A$ is also calculated for $^{10}$B + $^{10}$B, $^{16}$O + $^{16}$O, and $^{40}$Ca + $^{40}$Ca collisions at 0$-$10\% centrality at $\sqrt{s_{NN}}$ = 39 GeV, and the results are
presented in Fig.~\ref{B2-dif-PTA} (a) and (b). It is found that the coalescence parameter $B_{2}$ presents a system size dependence, i.e.
$B_{2}$ decreases as the system size increases.
This result is consistent with the centrality dependence in the same system such as Au + Au collisions. The  $p_{T}$ dependence of $B_{2}$ also presents an upward trend as shown in Fig.~\ref{B2}. The coalescence parameter $B_{3}$ is presented as a function of $p_{T}/A$ for the 0$-$10\% central collisions of
$^{10}$B + $^{10}$B, $^{16}$O + $^{16}$O, $^{40}$Ca + $^{40}$Ca, and  $^{197}$Au + $^{197}$Au systems at $\sqrt{s_{NN}}$ = 39 GeV in Fig.~\ref{B2-dif-PTA}(c), it shows the similar trend with $B_{2}$ even though the error remains larger.

\begin{figure}[htbp]
	\centering
	\includegraphics[width=8.7cm]{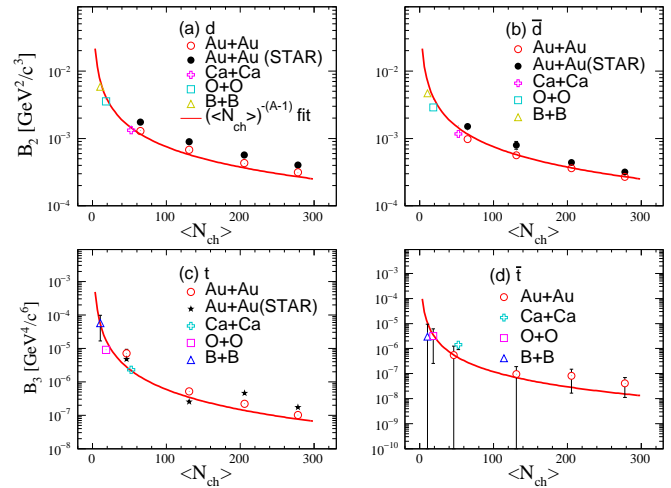}
	\caption {$\langle N_{ch} \rangle$ dependence of $B_{2}$ and $B_{3}$ for the (anti)deuteron and (anti)triton in $^{197}$Au + $^{197}$Au collisions at 0$-$10\%, 10$-$20\%, 20$-$40\%, and 40$-$60\% centralities as well as  the 0$-$10\% central collisions of  $^{10}$B + $^{10}$B, $^{16}$O + $^{16}$O and $^{40}$Ca + $^{40}$Ca systems at $\sqrt{s_{NN}} = 39$ GeV.
      The experimental data of (anti)deuteron produced in $^{197}$Au + $^{197}$Au collision  is taken from the STAR collaboration ~\cite{STARdeuteronPRC2019}.}
	\label{B2-dif-ntrack}
\end{figure}

Figure~\ref{B2-dif-ntrack} shows the $\langle N_{ch} \rangle$ dependence of coalescence parameters $B_{2}$ and $B_{3}$ of $d$ ($\bar{d}$) (a, b), $t$ ($\bar{t}$) (c, d) in $^{197}$Au+$^{197}$Au collisions at 0$-$10\%, 10$-$20\%, 20$-$40\%, and 40$-$60\% (40$-$80\% for $t$) centralities as well as  0$-$10\% central collisions of
 $^{10}$B + $^{10}$B, $^{16}$O + $^{16}$O, $^{40}$Ca + $^{40}$Ca systems at $\sqrt{s_{NN}} = 39$ GeV.
It is observed that the coalescence parameters $B_{2}$ and $B_{3}$ present an obvious collision centrality dependence, the values of $B_{2}$ for deuteron and anti-deuteron decrease with the increasing of  $\langle N_{ch} \rangle$. The $\langle N_{ch} \rangle$ dependence of $B_{3}$ for triton in $^{197}$Au + $^{197}$Au collisions at 0$-$10\%, 10$-$20\%, 20$-$40\%, and 40$-$80\% centralities at $\sqrt{s_{NN}} = 39$ GeV is also shown in this figure.  $B_{3}$ also shows a decreasing trend with  $\langle N_{ch} \rangle$. Besides, it is observed that the values of $B_{2}$ and $B_{3}$ present an obvious collision system dependence in 0-10\% central collisions of $^{10}$B + $^{10}$B, $^{16}$O + $^{16}$O, $^{40}$Ca + $^{40}$Ca, and $^{197}$Au + $^{197}$Au systems, the values of $B_{2}$ and $B_{3}$ for deuteron and anti-deuteron drop with the  increasing of  system size, the value of $B_{3}$ also shows a decreasing trend with  $\langle N_{ch} \rangle$.
Considering the properties of system size dependence from figure~\ref{fig:rRMS_if_Ntrack_Npart} as well as the relationship between coalescence parameter and nucleon correlation volume, i.e. $B_A \propto  1/V_{eff}^{A-1}$~\cite{CSERNAI1986223}, we found that $B_A$ can be expressed by a simple function, $B_A \propto 1/\left(\left<N_{ch}\right>\right)^{(A-1)}$ (here $A$ = 2 or  3). From the viewpoint of light nuclei production by coalescence mechanism, it is concluded that the coalescence parameter $B_A$ can reflect the collision system size when the system is at kinetic freeze-out stage.

 The thermal model has been successfully used to describe the multiplicities or particle ratios of hadrons and light nuclei~\cite{2018Natur.561..321A} in relativistic heavy-ion collisions, while the coalescence model basing on phase space data is another useful tool to treat light nuclei production. In practice, the phase-space data can be generated from various models, such as blast-wave model~\cite{PhysRevC.93.064909}, hydrodynamics~\cite{PhysRevC.102.044912}, transport model~%\cite{2020arXiv200500182S,PhysRevC.99.014901}~
 \cite{PhysRevC.99.014901} or pure analytical calculation~\cite{PhysRevC.95.044905}. In our work, the coalescence model basing on  the AMPT phase space data is used to study the light nuclei production at RHIC lower energy in the Beam Energy Scan project~\cite{bes1,bes2}, the results are consistent with the previous calculations and provide a more comprehensive understanding of the experiment data. Therefore we argue that these models could approach an equivalent simulation of the production of light nuclei assuming the thermal or kinetic freeze-out properties of the collision systems, respectively.

\section{Summary}
\label{summary}
In summary, based on the AMPT model coupled with the dynamic coalescence model, the collision system size dependence of light nuclei production was investigated for the 0$-$10\%, 10$-$20\%, 20$-$40\%, 40$-$60\%, and 60$-$80\% $^{197}$Au + $^{197}$Au collisions at $\sqrt{s_{NN}} = 39$ GeV. The calculated transverse momentum $p_T$ spectra can well describe the experimental data from the STAR collaboration and the extracted coalescence parameters of $B_{2}$ and $B_{3}$ fitted the data well. In the same way, the production of light nuclei is also calculated for the 0$-$10\% central collisions of $^{10}$B + $^{10}$B, $^{16}$O + $^{16}$O, and $^{40}$Ca + $^{40}$Ca systems at $\sqrt{s_{NN}} = 39$ GeV. As the system size is denoted by $\langle N_{ch} \rangle$ for different centralities and collision systems, the yields of light nuclei $\langle dN/dy \rangle$ present an obvious system size dependence, namely $\langle dN/dy \rangle$ increases with the system size ($\langle N_{ch} \rangle$). The reduction factor for light nuclei production is also presented for the system size dependence, which indicates that light nuclei production becomes more difficult in small systems. And the coalescence parameters $B_A$ ($A$ = 2, 3) as a function of $\langle N_{ch} \rangle$ fall into the same group regardless for different centralities in a fixed collision system or different systems at a fixed centrality. Coalescence  parameters $B_A$ ($A$ = 2, 3) present a decreasing trend with the increasing of $\langle N_{ch} \rangle$, i.e. follow  a proportional dependence on  $1/\langle N_{ch} \rangle^{A-1}$.  We can conclude that the light nucleus  production essentially depends on the fireball volume, reflected in the system size or centralities. These results shed light on further experimental system scan project at RHIC or LHC.

\begin{acknowledgements}

This work was supported in part by the National Natural Science Foundation of China under contract Nos. 11875066, 11890710, 11890714, 11925502, 11961141003, National Key R\&D Program of China under Grant No. 2018YFE0104600 and 2016YFE0100900, the Strategic Priority Research Program of CAS under Grant No. XDB34000000, the Key Research Program of Frontier Sciences of the CAS under Grant No. QYZDJ-SSW- SLH002, and the Guangdong Major Project of Basic and Applied Basic Research No. 2020B0301030008.

\end{acknowledgements}

\end{CJK*}

\bibliography{mybibfile}

\end{document}